\documentclass[12pt]{article}
\usepackage{graphicx}
\usepackage{amsmath}
\usepackage{hyperref}
\hypersetup{colorlinks=false}

\newcommand\pubnumber{}
\newcommand\pubdate{\today}

\newcommand{\LSB}{\raisebox{-0.3ex}{\mbox{\LARGE$\left[\right.$}}}
\newcommand{\RSB}{\raisebox{-0.3ex}{\mbox{\LARGE$\left.\right]$}}}

\def\roma{Physics Department \\
University of Rome ``La Sapienza" \\ piazz.le Aldo Moro 2, I-00185 Rome, Italy }

\def\Title#1{\begin{center} {\Large #1 } \end{center}}
\def\Author#1{\begin{center}{ \sc #1} \end{center}}
\def\Address#1{\begin{center}{ \it #1} \end{center}}

\newcommand\pubblock{\rightline{\begin{tabular}{l} \pubnumber\\
         \pubdate  \end{tabular}}}
\newenvironment{Abstract}{\begin{quotation}  }{\end{quotation}}
\newenvironment{Presented}{\begin{quotation} \begin{center} 
             PRESENTED AT\end{center}\bigskip 
      \begin{center}\begin{large}}{\end{large}\end{center} \end{quotation}}
\def\Acknowledgements{\bigskip  \bigskip \begin{center} \begin{large}
             \bf ACKNOWLEDGEMENTS \end{large}\end{center}}

\def\beq{\begin{equation}}
\def\eeq#1{\label{#1}\end{equation}}
\def\eeqn{\end{equation}}

\def\beqa{\begin{eqnarray}}
\def\eeqa#1{\label{#1}\end{eqnarray}}
\def\eeqan{\end{eqnarray}}

\let\bar=\overbar

\def\Dslash{\not{\hbox{\kern-4pt $D$}}}
\def\dslash{\not{\hbox{\kern-2pt $\del$}}}

\def\msb{{\bar{\ssstyle M \kern -1pt S}}}

\begin{document}
\begin{titlepage}
\pubblock

\vfill
\Title{$\rm{K}^0-\bar{\rm{K}}^0$ on the Lattice}
\vfill
\Author{ Petros Dimopoulos}
\Address{\roma}
\vfill
\begin{Abstract}
\noindent I review  recent lattice calculations performed with $N_f=2$ and $N_f=2+1$ dynamical fermions  
which provide a precise computation of the $B_{\rm{K}}$ bag parameter. I also report on 
$N_f=2$ dynamical quark simulations aiming at the computation 
of the full basis of the $\Delta S=2$ four-fermion operator matrix elements 
that are relevant to models beyond the Standard Model.   
\end{Abstract}
\vfill
\begin{Presented}
Proceedings of CKM2010, \\ the 6th International Workshop \\ on the CKM Unitarity Triangle,  \\
University of Warwick, UK, 6-10 September 2010
\end{Presented}
\vfill
\end{titlepage}
\def\thefootnote{\fnsymbol{footnote}}
\setcounter{footnote}{0}

\section{Introduction}
The indirect CP violation, present in the neutral Kaon meson pair ($\rm{K}^0-\bar{\rm{K}}^0$) oscillations 
\cite{Christenson:1964}, 
is a confirmed experimental fact. The neutral Kaons ${\rm K}^{0}=(\bar{s}d)$ and $\bar{\rm{K}}^{0}=(s\bar{d})$ 
are flavour eigenstates which, in the Standard Model (SM), can mix to lowest order through a box diagram 
due to the weak interactions. 
The weak hamiltonian describing the time evolution of the kaon pair system can be expressed via 
a 2 $\times$ 2 non-hermitian matrix,
$H=M-i \Gamma/2$. The dispersive and absorptive parts,  $M$ and $\Gamma$ respectively, are hermitian matrices.
Defining the CP eigenstates 
$|K_{\pm}\rangle = (|K^{0} \rangle \pm |\bar{K}^{0} \rangle)/\sqrt{2}$, 
it turns out that the two hamiltonian eigenstates, called $K-$\,short and $K-$\,long, are  
$|K_{\scriptstyle{S/L}} \rangle = \frac{\textstyle{1}}
{\textstyle{\sqrt{1+\bar{\epsilon}^2}}}~(|K_{\pm} \rangle + \bar{\epsilon} |K_{\mp} \rangle)$.
The paramater $\bar{\epsilon}$ represents the mixing between the two CP eigenstates; it depends on the 
phase conventions of the eigenstates and represents 
the theoretical measure of the indirect CP violation. On the other hand the experimental measurement involves a phase 
independent quantity which is
\begin{equation}
\epsilon_{\rm{K}} = \frac{{\cal A}(K_{L} \longrightarrow (\pi \pi)_{I=0}))}
{{\cal A}(K_{S} \longrightarrow (\pi \pi)_{I=0}))}  
\end{equation}
It turns out that (Ref.~\cite{chau}) $\epsilon_{\rm{K}} = \bar{\epsilon} + i \xi$, where 
$\xi=\frac{\textstyle{\rm{Im}\, A_0}}{\textstyle{\rm{Re}\, A_0}}$ and $A_0$ is the  isospin zero amplitude 
for the  $\rm{K}^0$ decay to either $(\pi^0 \pi^0)$ or $(\pi^- \pi^+)$.  
It has been shown (~\cite{BurGuad},\cite{BurGuadIsi}, \cite{LenzNier}; see also \cite{Guad_CKM}) that, 
under reasonable estimates for the long-distance contribution to both the dispersive and the absorptive  
parts of the hamiltonian, $\epsilon_{\rm{K}}$ is approximately given 
\begin{equation} \label{eps_K}
 \epsilon_{\rm{K}} \simeq \kappa_{\epsilon} \frac{e^{i \phi_{\epsilon}}}
{\sqrt{2}} \frac{\rm{Im}\, M_{12}}{\Delta m_{\rm{K}}} 
\end{equation}
In Eq.~(\ref{eps_K})   the phase $\phi_{\epsilon}= 43.51(5)^o$ and the mass difference
between $\rm{K}_{L}$ and $\rm{K}_{L}$,  $\Delta m_{\rm{K}}=3.483(6) \times 10^{-12}~\rm{MeV}$, 
have been 
determined experimentally  with high precision~\cite{Nakamura}. 
The factor $\kappa_{\epsilon}=0.94(2)$  incorporates in an approximate way  
the effect of the long distance contributions mentioned above.

In the SM the calculation of $\rm{Im}\, M_{12}$ is performed in the Operator Product Expansion (OPE)
approach. Once the heavy degrees of freedom, including the charm quark, are integrated out the CP violation contribution in the neutral K-meson mixing is described in terms of the $\bar{K}^0-K^0$ matrix element of a local  $\Delta S=2$ 
four-fermion operator:
\begin{equation}\label{M12}
 M_{12} = \frac{1}{2 m_{\rm{K}}} (\langle \bar{K}^{0}|{\cal H}_{eff}^{\Delta S=2}|K^{0}\rangle)^{*} =
\frac{1}{2 m_{\rm{K}}} C_{{\cal W}}^{\rm{SM}}(\mu) (\langle \bar{K}^{0}|{\cal O}^{\Delta S=2}(\mu)|K^{0}\rangle)^{*}
\end{equation}
The Wilson coefficient $C_{{\cal W}}^{\rm{SM}}(\mu)$ contains all the short distance effects which are 
calculated in perturbation theory.   
The matrix element of the four-fermion operator ${\cal O}^{\Delta S=2} \equiv {\cal O}_1=(\bar{s}d)_{V-A}(\bar{s}d)_{V-A}  
\equiv \bar{s}\gamma_{\mu}(1-\gamma_5)d\,\, \bar{s}\gamma_{\mu}(1-\gamma_5)d$ 
is written in the form 
\begin{equation}\label{BK}
 \langle \bar{K}^{0}|{\cal O}_1(\mu)|K^{0}\rangle = \frac{8}{3} m_{\rm{K}}^2 f_{\rm{K}}^2 B_{\rm{K}}(\mu)
\end{equation}
where the parameter $B_{\rm{K}}$ is the amount by which it differs from its Vacuum Saturation Approximation (VSA)
value  and 
$m_{\rm{K}}$ and $f_{\rm{K}}$ are the mass and the decay constant of the  Kaon respectively.
Therefore the bag parameter $B_{\rm{K}}$ is the measure of the non-perturbative QCD contribution   
in the hadronic matrix element of the K-meson pair oscillation.
The Renormalisation Group Invariant (RGI) quantity is defined by the equation
\begin{equation}\label{BKrgi}
 \hat{B}_{\rm{K}}=\LSB \alpha_{s}^{(3)}(\mu) \RSB^{-2/9}~\LSB 1 + \frac{\alpha_{s}^{(3)}(\mu)}{4\pi}\,J_3 \RSB
B_{\rm{K}}({\mu})
\end{equation}
where $\alpha_{s}^{(3)}(\mu)$ is the running coupling constant for  $N_f=3$ flavours and $J_3$ has been computed 
up to  NLO order \cite{Ciuchini:1997bw}.
Eqs.~(\ref{eps_K})-(\ref{BKrgi}) yield
\begin{equation}\label{eps_final}
 |\epsilon_{K}| = \kappa_{\epsilon} C_{\epsilon} 
\hat{B}_{\rm{K}}\, |V_{cb}|^2 \lambda^2 \bar{\eta} 
         \LSB-\eta_1 S_0(x_c) (1-\frac{\lambda^2}{2}) + \eta_{2} S_{0}(x_t) |V_{cb}|^2 \lambda^2 (1-\bar{\rho})
          + \eta_{3} S_{0}(x_c,x_t) \RSB   
\end{equation}
where $C_{\epsilon}=\frac{G_F^2 f_{\rm{K}}^2 m_{\rm{K}}M_{\rm{W}}^2}{6\sqrt{2}\pi^2 \Delta m_{\rm{K}}}$ and 
$\lambda=|V_{us}|$. $S_0$ are the Inami-Lim functions depending on 
$x_{c,t}=m_{c,t}^2/M_{\rm{W}}^2$ and give the charm, top and 
charm-top contributions to the box diagram; $\eta_{i=1,2,3}$ contain  the corresponding 
short distance QCD contributions to NLO order\footnote{See Ref.~\cite{Brod} on a recent calculation for $\eta_3$ 
to  NNLO order.} (\cite{Buras1990}-\cite{Herrlich1996}).    
The experimental value of $|\epsilon_{K}|=2.228(11)\times 10^{-3}$ is known with high 
precision \cite{Nakamura}. Therefore, the hyperbola 
defined by Eq.~(\ref{eps_final}) in the  $(\bar{\rho}, \bar{\eta})$ plane can be used to  (over-)constrain the upper 
vertex of the unitarity triangle with a precision that depends on the quality of the 
estimates of $|V_{us}|$, $|V_{cb}|$ and 
$\hat{B}_{\rm{K}}$. While the numerical value of the first is known to very good precision (see Ref.~\cite{FLAG}), 
the estimate of (inclusive) $|V_{cb}|$ (Ref.~\cite{HFAG}) is still given with un uncertainty 
of $\sim 2\%$ which gets amplified  four times  in 
Eq.~(\ref{eps_final}) because  it enters to the fourth power.  
$\hat{B}_{\rm{K}}$ can be calculated on the lattice from first principles.  
Its estimate, until 2008, used to represent the largest source of uncertainty in 
Eq.~({\ref{eps_final}) but now, thanks to   recent precision lattice calculations, it is known with
a total (statistical and systematic) uncertainty of less than  $4\%$. 
This is the consequence of employing unquenched simulations with $N_f=2$ and $N_f=2+1$ dynamical quarks with high 
statistics, controlling with better accuracy  both the discretization errors  and the extrapolation 
to the physical point and using non-perturbative methods for the renormalisation of the operators on the lattice.

\section{$\mathbf{B_{\rm{K}}}$ calculation on the lattice}      
\subsection{General considerations}

 The evaluation of $B_{\rm{K}}$ on the lattice requires the computation 
of a three point correlation function with the insertion of two pseudoscalar  meson interpolating fields  
at two  time slices, say, $t_1$ and $t_2$ and the insertion of the $\Delta S=2$ four-fermion  operator 
considered at any 
time slice $t_0$ with $t_0 \in [t_1, t_2]$. In order to  carry out the computation of  
 Eq.~(\ref{BK}), the three point correlation function needs to be divided by two two-point correlation functions 
 each of which  involves a pseudoscalar meson  and an axial current interpolating fields;  then  
 one takes the asymptotic limit i.e. $t_1 \ll t_0 \ll t_2$ in order to obtain the value of the matrix element 
 between the lowest-lying  states. In summary,  one has
 \begin{equation}\label{BK_calc}
\frac{\langle {\cal P}(t_{1})^{\dagger} \, {\cal O}^{\Delta S=2}(t_0) \, {\cal P}(t_{2}) \rangle}
         {(8/3){\langle \cal P}(t_{1})^{\dagger} {\cal A}_0(t_0) \rangle \,
       \langle {\cal A}_0(t_0)^{\dagger} {\cal P}(t_{2}) \rangle} 
       \xrightarrow{t_1 \ll t_0 \ll t_2}
 \frac{\langle{\bar{K}^0}|{\cal O}^{\Delta S=2}|{K^0} \rangle}
            {(8/3) \langle \bar{K}^0|A_{0}|0 \rangle \langle 0|A_{0}^{\dagger}|K^{0}\rangle} = B_{\rm{K}}             
 \end{equation}
where ${\cal P}(t)$ and ${\cal A}(t)$ denote the pseudoscalar density and axial current interpolating 
fields respectively. Note also that 
  $\langle 0|A_{0}|K^{0}\rangle = f_{\rm{K}} \, m_{\rm{K}}$. In practice, due to parity conservation 
in  QCD, one needs  to evaluate only the matrix element of the parity-even part, 
${\cal O}_{VV+AA}=\bar{s}\gamma_{\mu}d \bar{s}\gamma_{\mu}d +
\bar{s}\gamma_{\mu}\gamma_5d \bar{s}\gamma_{\mu}\gamma_5d$, of the ${\cal O}^{\Delta S=2}$ operator. 

Naturally,  a renormalization step is necessary to get finite results in the continuum limit ($a \rightarrow 0$).
For this purpose the renormalization   
constant (RC) of the operator ${\cal O}$, $Z_{{\cal O}}$, (or a full matrix of RCs in the general case where
 more than one operators get mixed in the renormalisation procedure) has to be computed at a certain 
scale $\mu$ (the same as that  of the Wilson coefficient). 
Lattice perturbation theory can be used. The result, however, suffers from more or 
less insufficiently well estimated systematic errors due to the truncation of the perturbative expansion. 
This is  especially true for not so fine  lattice spacings. 
In any case non-perturbative methods offer a more accurate way in computing  the RC of 
operators.  By imposing renormalization conditions on correlation 
functions with operator insertions and 
performing calculations of matrix elements on the lattice, one has the opportunity to embody  
in the calculation  both higher order contibutions and non-perturbative effects. 
In this way the remaining sources of systematic error are only due to the use of 
 perturbation theory in passing from the lattice scheme 
(at some scale $\mu$) to a continuum renormalisation scheme.  

Lattice calculations  are performed at a number of various values of lattice spacing and suffer from discretization 
errors which  can be  eliminated when the extrapolation to the continuum limit (c.l.) is  achieved. 
This source of systematic error 
can be well kept under control if the values of the lattice spacing      
are small, typically  less than about 0.1 fm.  Then  simulations for at least three values of the lattice spacing are 
typically necessary to be able to perform the extrapolation to the continuum limit. In most of the cases 
discretization error of the various 
physical quantities measured on the lattice  are of  at most $O(a^2)$.  

It is interesting to have a look at the ``world-average'' values of  
$\hat{B}_{\rm{K}}$ which cover a period of the past fifteen years as they reported at the Lattice Conferences. 
The RGI results  (computed with the same running using $N_f=3$ flavours, see  Ref.~\cite{Lublat2009}) are shown 
in Fig.~\ref{fig:BK_wav}.  
 

\begin{figure}[htb]
\vspace*{-0.2cm}
\centering
\includegraphics[height=2.0in]{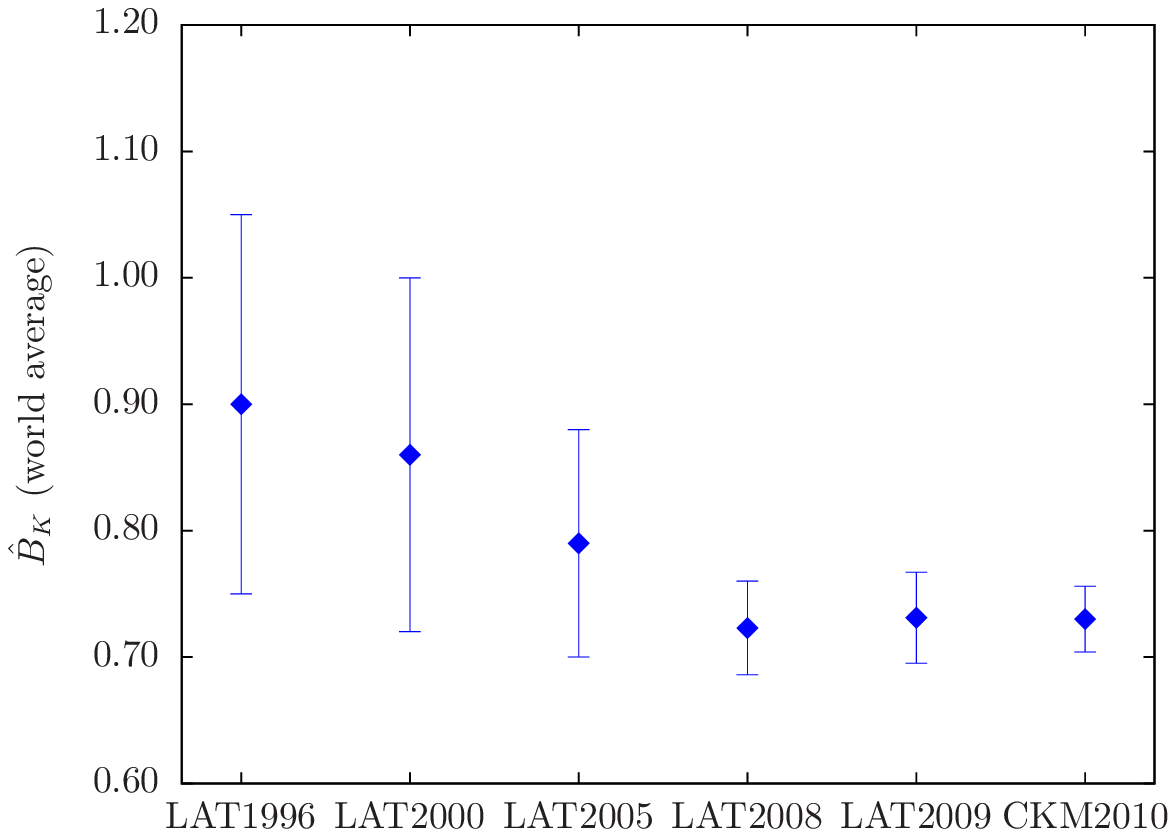}
\caption{$\hat{B}_{\rm{K}}$ ``world averages'' over the last fifteen years reported at the Lattice 
Conferences. The world average result at CKM2010 is also included.}
\label{fig:BK_wav}
\end{figure}
It is impressive to notice the great improvement in the quality of the results where 
total uncertainty  has decreased 
by almost five times in the last fifteen years. It should also be noticed that even until 2005 the $\hat{B}_{\rm{K}}$
``world average'' result was based on estimates computed in the quenched approximation
 with the systematic error due to 
quenching being in practice unknown. However, the ``world average'' value of the last 
two years is based only on
results produced with unquenched simulations. 
  
\subsection{Recent lattice $\mathbf{B_{\rm{K}}}$ computations}

Different lattice groups compute $B_{\rm{K}}$  using a variety of 
lattice regularizations which conserve (most of) the chiral symmetry and provide well controlled and small $O(a^2)$ 
discretization errors. Besides that, simulations 
have been performed at a well tuned physical value for the strange quark mass while the lightest simulated quark mass
 value, even if
not yet at  its physical value,  has smaller values than in any preceding calculation of $B_{\rm{K}}$.
Four collaborations, performing unquenched simulations, have computed the continuum limit value for 
$B_{\rm{K}}$. Three of them have simulated $N_f=2+1$ dynamical quarks, while the fourth one has used $N_f=2$
dynamical quark simulations.   

\begin{table}[!h]
\vspace*{-1.5cm}
\hspace*{-1.5cm}
\scalebox{0.85}{
\begin{tabular}{llccccc}
\hline \hline
&&&&&& \\
Collaboration &  $N_f$    & fermion discr. & $a \,[\rm{fm}]$   & $\LSB M_{\pi}^{\rm{val}}/M_{\pi}^{\rm{sea}} \RSB_{\rm{min}}$    
&  $\LSB M_{\pi}^{\rm{val}} L \RSB_{\rm{min}}$ & Renorm/tion    \\
      &           &  val. / sea    &                   &          \hspace*{-0.5cm}in [\rm{MeV}] & &   \\           
\hline 
&&&&&& \\
ALV \hspace*{1.74cm}\cite{ALV} & 2+1 & DW/Asqtad & \hspace*{-1.cm}0.12, 0.09             & 240/370 & 3.5 & Non-Pert.  \\
BSW \hspace*{1.642cm}\cite{BSW} & 2+1 & HYP-stag/Asqtad & 0.12, 0.09, 0.06 & 240/300 & 2.5 & Pert.  \\
RBC-UKQCD \cite{RBC} & 2+1 & DW/DW & \hspace*{-1.cm}0.11, 0.09& 220/290 & 3.1 & Non-Pert.  \\
&&&&&& \\
\hline
&&&&&& \\
ETMC \hspace*{1.33cm}\cite{ETMC} &  2 & OS/TM & 0.10, 0.09, 0.07 & 280/280 & 3.3 & Non-Pert.  \\
\hline\hline
\end{tabular}
}
\caption{Simulation details of unquenched simulations of $B_{\rm{K}}$ computation with  $N_f=2$ and $N_f=2+1$ 
dynamical quarks.
 }
\label{Table:BKres1}
\end{table}

Table~\ref{Table:BKres1}  gives a summary of the simulation details for the computation 
by each collaboration; Table~\ref{Table:BKres2} contains $B_{\rm{K}}$ results 
expressed  in the $\bar{\rm{MS}}$ scheme at 2 GeV as well as in the RGI definition 
(see Refs. \cite{ALV}-\cite{ETMC}); the first error is  statistical while the second is 
the systematic one; the total error has to be
calculated as the sum in quadrature of the statistical and the systematic ones.
In the following paragraphs a short description of each computation is presented.

Aubin, Laiho and Van de Water (ALV) (see Ref.~\cite{ALV}) have presented a mixed action calculation 
performed with Domain Wall (DW) fermion regularization in the valence quark sector 
on $N_f=2+1$ dynamical Asqtad-improved  staggered quarks (MILC). 
DW fermions in the valence sector offer the possibility to use the MILC dynamical configurations without having
to bother about  
taste mixing. Since they  have  a residual chiral breaking 
(due to the finite size of the fifth lattice dimension), the renormalisation of the 
parity even four-fermion operator  suffers from  mixing with operators of 
wrong chirality through coefficients  whose values are expected to be suppressed at order $O((am_{res})^2)$. 
The implementation of the non-perturbative renormalisation method  
(RI-MOM) is straightforward for DW fermions.  
They have done simulations at two values of
lattice spacing, namely $a=0.12$ and $0.09$ fm. 
They extrapolate to the physical point using fit functions based on Mixed action $\chi$PT 
with the addition of NNLO analytical terms.  Their result is given in Table~\ref{Table:BKres2} ; 
a large part of the systematic error quoted is due to the uncertainty in the renormalisation constant computation.

Bae {\it et al.} (BSW\footnote{It is an acronym for Brookheaven, Seoul and Washington groups.}, 
see Ref.~\cite{BSW}) have also completed a mixed action computation where they have used 
HYP-smeared staggered valence 
quarks on MILC $2+1$ dynamical quark configurations. 
The choice for this particular valence quark regularization is 
justified by its property of dispalying rather small taste breaking effects 
and a computationally cheap implementation. They extrapolate to the physical point trying 
fit functions based on SU(3) and SU(2) Staggered-$\chi$PT. The two fitting procedures 
give compatible results with the latter being more straightforward in fitting the data. 
Simulations have been carried out  at three values of lattice 
spacing, $a=0.12$, 0.09 and 0.06 fm. The dominant source of uncertainty is due to the perturbative 
approach they use to  renormalize  the four-fermion operator. In fact, the  total error 
is almost $6 \%$ the larger  part of which is due to  RC's uncertainty.    

RBC-UKQCD Collaboration (Ref.~\cite{RBC}) used a $N_f=2+1$ DW action both for the sea and 
the valence quarks at two values
of the lattice spacing, $a=0.11$ and 0.09 fm. The  renormalization of the four-fermion operator 
is performed employing
RI-MOM methods where  non-exceptional momentum renormalization conditions and twisted boundary conditions
have been applied. They achieve  further suppression 
of the residual wrong chirality mixing and a matching to $\bar{\rm{MS}}$ at a higher scale value (3 GeV) which
allows for a a better control of the perturbative systematic uncertainty. The uncertainty to the final value 
due to the renormalisation procedure is estimated 2$\%$.    
The physical point is reached via a combined continuum and SU(2) PQ$\chi$PT 
fitting formula. The total error in the final value  is  $3.6\%$, while the pure stastistical one  is  
about $0.9\%$.

ETM Collaboration (Ref.~\cite{ETMC}) has adopted a mixed action set-up  to perform the computation with 
a Wilson fermion regularization. Actually, since Wilson fermions  explicitly break the 
chiral symmetry, the relevant $B_{\rm{K}}$ computation done with the use of plain Wilson quarks 
suffers from  $O(a)$ discretization errors and yields  wrong chirality mixing in 
the process of renornalizing  the lattice ${\cal O}_{VV+AA}$ operator. 
It is known that both problems can inject serious systematic uncertainties in the computation. 
However, it has been shown 
that they can be tackled simultaneously by employing Osterwalder-Seiler (OS) valence  on   
Twisted Mass (TM) sea fermions, both tuned at maximal twist (see Ref.~\cite{FR2}). ETMC has carried out the 
computation working with $N_f=2$ dynamical quark simulations, so the strange quark is still  quenched. 
Unitarity violations due to the mixed action set-up, that are expected to be $O(a^2)$ effects,  have been 
shown to be well under control in the continuum limit. 
The extrapolation to the physical light quark mass is performed 
using a fit function based on SU(2)-$\chi$PT while three values of the lattice spacing, 
$a = 0.10$, 0.09 and 0.07 fm,  have been used to extrapolate to the continuum limit. 
The renormalization of the four-fermion operator
has been carried out in a non-perturbative way using the RI-MOM method. The mixing with operators of wrong chirality 
has been shown to be numerically negligible.  The total error  is  $4 \%$.    

\begin{table}[!t]
\vspace*{-1.5cm}
\hspace*{2cm}
\begin{tabular}{lcccc}
\hline \hline
&&&& \\
Collaboration &               & $N_f$    &  $B_{\rm{K}}^{\bar{\rm{MS}}}( 2 \, \rm{GeV})$ & $\hat{B}_{\rm{K}}$   \\
&&&& \\
\hline 
&&&& \\
ALV       & \cite{ALV} & 2+1 & 0.527(06)(20) & 0.724(08)(28) \\
BSW       & \cite{BSW} & 2+1 & 0.529(09)(32) & 0.724(12)(43) \\
RBC-UKQCD & \cite{RBC} & 2+1 & 0.549(05)(26) & 0.749(07)(26) \\
&&&& \\
\hline
&&&& \\
ETMC      & \cite{ETMC} & 2  & 0.517(18)(11) & 0.729(25)(17) \\
\hline\hline
\end{tabular}
\caption{Results  in the continuum limit for the  $B_{\rm{K}}$-parameter expressed in the $\bar{\rm{MS}}$ scheme 
and in the RGI definition from unquenched simulations with $N_f=2$ and $2+1$ dynamical quarks. 
The first error is quoted as statistical
while the second is the estimate of  the systematic uncertainties.}
\label{Table:BKres2}
\end{table}

The $\hat{B}_{\rm{K}}$ results published in the last five years and computed on quenched ($N_f=0$) 
and unquenched lattices 
with $N_f=2$ and $N_f=2+1$ dynamical quark simulations are collected in Fig.~\ref{fig:BK_results}. 
The label ({\it c.l.}) is attached to the  $\hat{B}_{\rm{K}}$ values obtained after the continuum limit has been 
taken.
 Vertical lines 
indicate the average $\hat{B}_{\rm{K}}$ result with its error, based on $N_f=2+1$ simulations, which is
\begin{equation}\label{2+1}
 \hat{B}_{\rm{K}} = 0.736(05)(26)  ~~~~ @ ~~~~ N_f=2+1
\end{equation}
This value has been obtained by taking the  average of the $N_f=2+1$ results weighted  with  
the quoted statistical error. The systematic uncertainty is the smallest error quoted in Table~\ref{Table:BKres2}.
The total uncertainty is about $3.6\%$. 
The $\hat{B}_{\rm{K}}$ average  for  the $N_f=2$ case obviously coincides with the result provided by ETMC, 
since it is the 
only continuum limit result available. One finds 
\begin{equation}\label{2}
 \hspace*{-0.9cm}\hat{B}_{\rm{K}} = 0.729(25)(16)  ~~~~ @ ~~~~ N_f=2
\end{equation}   
Results from Eqs.~(\ref{2+1}) and (\ref{2})  indicate that the systematic uncertainty due to the quenching of the 
strange quark has an impact  which is  smaller than other systematic errors. 
Moreover, comparing the result of Eq.~(\ref{2+1}) with the most 
precise of the quenched results (see Ref.~\cite{CPPACS}), one can infer that the systematic uncertainty due to the 
quenching is small (actually $\sim  6\%$).

\begin{figure}[htb]
\centering
\includegraphics[height=3.0in]{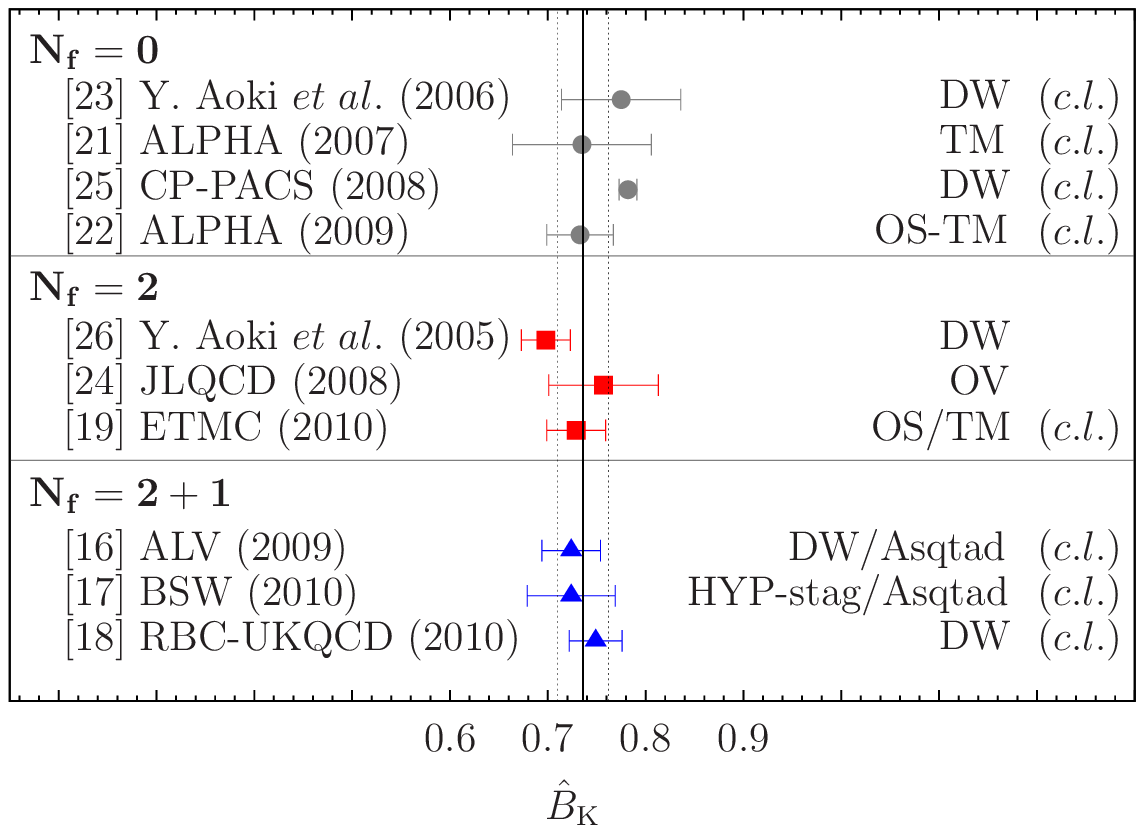}
\caption{$\hat{B}_{\rm{K}}$ results from $N_f$ = 0 (quenched) and unquenched simulations with $N_f$ = 2 and $N_f$ = 2+1 
dynamical quarks. Computations where  the continuum limit has been taken are indicated with ``({\it c.l.})''. 
Vertical lines
refer to the average result of Eq.~(\ref{2+1}). In the plot, results from pioneer works  
 with $N_f$ = 2 \cite{Flynn} and $N_f$ = 2+1 \cite{Gamiz} unquenched simulations have not been included 
since their quoted total uncertainty is now considered too large compared to today standards. }
\label{fig:BK_results}
\end{figure}

\section{ $\mathbf{K^0-\bar{K} ^0}$mixing beyond the SM}

There are various models for the physics  Beyond the SM (BSM)   
which lead to other possible $\Delta S=2$ processes 
at one loop. In this case the computation of the relevant matrix elements of the effective hamiltonian 
in combination with the experimental value of $\epsilon_{\rm{K}}$ would offer 
the chance of obtaining constraints  on the parameters of the model  (like for instance estimates on the 
off-diagonal terms of the squark matrix in  supersymmetric models \cite{Ciuchini-etal98}) 
which enter explicitly in the Wilson coefficients. 

The general form for the $\Delta S=2$ effective Hamiltonian in  BSM models reads  

\begin{equation}\label{Heff_gen}
{\cal H}_{\rm{eff}}^{\Delta S=2} = \sum_{i=1}^{5} C_i(\mu) {\cal O}_i 
+ \sum_{i=1}^{3} \tilde{C}_i(\mu) \tilde{{\cal O}}_i 
\end{equation}
where 
\begin{eqnarray}\label{O_basis}
{\cal O}_1 &=& [\bar{s}^a \gamma_\mu (1-\gamma_5)d^a][\bar{s}^b \gamma_\mu (1-\gamma_5)d^b], \nonumber \\
{\cal O}_2 &=& [\bar{s}^a (1-\gamma_5)d^a][\bar{s}^b  (1-\gamma_5)d^b], ~~~~~~~~
{\cal O}_3 = [\bar{s}^a (1-\gamma_5)d^b][\bar{s}^b  (1-\gamma_5)d^a], \nonumber \\
{\cal O}_4 &=& [\bar{s}^a (1-\gamma_5)d^a][\bar{s}^b  (1+\gamma_5)d^b], ~~~~~~~~
{\cal O}_5 = [\bar{s}^a (1-\gamma_5)d^b][\bar{s}^b  (1+\gamma_5)d^a] \nonumber \\ 
\tilde{{\cal O}}_1 &=& [\bar{s}^a \gamma_\mu (1+\gamma_5)d^a][\bar{s}^b \gamma_\mu (1+\gamma_5)d^b], \nonumber \\
\tilde{{\cal O}}_2 &=& [\bar{s}^a (1+\gamma_5)d^a][\bar{s}^b  (1+\gamma_5)d^b], ~~~~~~~~
\tilde{{\cal O}}_3 = [\bar{s}^a (1+\gamma_5)d^b][\bar{s}^b  (1+\gamma_5)d^a]  
\end{eqnarray}
We have seen that in the SM case, Eq.~(\ref{M12}),  only the operator ${\cal O}_1$  contributes. \\
The  parity-even parts of the operators $\tilde{{\cal O}}_i$
coincide with those of the operators ${\cal O}_i$. Therefore, due to parity conservation 
in the strong interactions only the parity-even contribution of the operators ${\cal O}_i$ need to be calculated.  
A basis of the parity-even operators is
\begin{eqnarray}\label{O_even}
O^{VV} &=& (\bar{s}\gamma_{\mu}d)(\bar{s}\gamma_{\mu}d), ~~~~~~~~ 
O^{AA} = (\bar{s}\gamma_{\mu}\gamma_5 d)(\bar{s}\gamma_{\mu}\gamma_5 d), \nonumber \\
O^{PP} &=& (\bar{s}\gamma_5 d)(\bar{s}\gamma_5 d), ~~~~~~~~~
O^{SS} = (\bar{s}d)(\bar{s}d), \nonumber \\
O^{TT} &=& (\bar{s}\sigma_{\mu \nu} d)(\bar{s} \sigma_{\mu \nu} d) 
\end{eqnarray}
in terms of which and after using  a Fierz transformation one obtains the relations
\begin{eqnarray} \label{BSM_op}
{\cal O}_1 &=& (O^{VV} + O^{AA}), ~~~~ 
{\cal O}_2 = (O^{SS} + O^{PP}),   ~~~~ 
{\cal O}_3 = -\frac{1}{2}(O^{SS} + O^{PP} - O^{TT}),  \nonumber \\
{\cal O}_4 &=& (O^{SS} - O^{PP}), ~~~~~
{\cal O}_5 = -\frac{1}{2} (O^{VV} - O^{AA})   
\end{eqnarray}

\noindent The B-parameters for the operators of Eq.~(\ref{BSM_op}) are defined as 

\begin{eqnarray}
 \langle \bar{K}^{0} | {\cal O}_1(\mu) | K^{0} \rangle &=& B_1(\mu) \frac{8}{3} m_{K}^{2} f_{K}^{2}  \equiv B_K(\mu)
  \frac{8}{3} m_K^2 f_K^2 \nonumber \\
  \langle \bar{K}^{0} | {\cal O}_i(\mu) | K^{0} \rangle &=&  C_i B_i(\mu) 
\LSB \frac{ m_{K}^{2} f_{K}}{ m_s (\mu) + m_d(\mu)} \RSB^2, \nonumber 
\end{eqnarray}
where $C_i = \{-5/3, 1/3, 2, 2/3\}, ~~ i=2, \ldots, 5$.
The matrix element of the operator ${\cal O}_1$ vanishes in the chiral limit while the matrix element of the 
operators ${\cal O}_i,~ i=2,\ldots,5$ get a  non-zero value in the chiral limit. From the above equations 
it can be seen that the calculation of the $B_i$ parameters for $i=2,\ldots,5$ involves the calculation of the 
quark mass at a common renormalization scale $\mu$.  In order to avoid any extra systematic uncertainties 
in the computation of  the matrix elements due to the quark mass
evaluation,  an alternative calculation  has been proposed   which consists in calculating directly 
appropriate ratios of the ${\cal O}_{i}, ~i=2, \ldots, 5$ four-fermion matrix elements 
with the ${\cal O}_1$ one\footnote{Note that in forming these ratios one should take care of the fact that 
the matrix element in the denominator vanishes in the chiral limit.} 
(\cite{Donini:1999nn}, \cite{Babich_etal06}). 

There are a few quenched lattice calculations performed with the use of tree-level improved Wilson fermions 
\cite{Donini:1999nn}, overlap fermions \cite{Babich_etal06} and DW fermions \cite{Nakamura_etal06}.
However only the first two computations have been carried out using two values of the lattice spacing 
allowing for the possibility of estimating  discretisation effects.
A preliminary study referring to the bare matrix elements computation using $N_f=2+1$ dynamical DW fermions was also
presented in Ref.~\cite{Wennekers:2008sg}.    
Recently ETMC has presented a $N_f=2$ unquenched calculation using the mixed action set-up  described above (Section 
2.2). Both $B_{i}$-bag parameters and matrix elements ratios have been calculated at three values of the 
lattice spacing, $a=0.1$, 0.09 and 0.07 fm from which  reliable continuum limit estimates can be obtained.  
The results presented in Ref.~\cite{ETMC_B_BSM} are  {\it preliminary}; however, it might be useful to 
proceed at a first comparison between the quenched and the unquenched estimates for  the bag-parameters
$B_i, ~i=2, \ldots, 5$ calculated in the $\bar{\rm{MS}}$ scheme at 2 GeV. This comparison is provided  
in Fig.~\ref{fig:Bi_comp}.             
 
\begin{figure}[htb]
\centering
\vspace*{-3cm}
\includegraphics[height=5.0in, angle=-90]{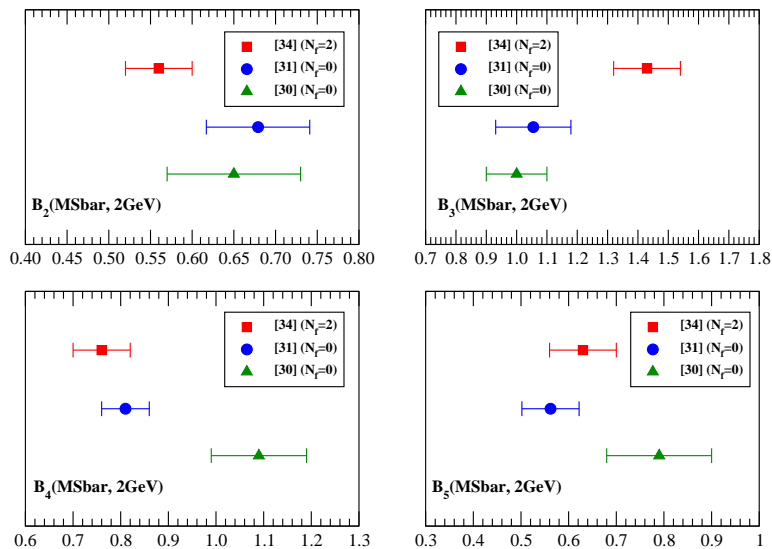}
\vspace*{-1.1cm}
\caption{Comparison of continuum limit $B_i, ~(i=2, \ldots, 5)$ results (in $\bar{\rm{MS}}$ scheme at 2 GeV)  
obtained using $N_f$ = 0 (quenched) (\cite{Donini:1999nn}, \cite{Babich_etal06}) 
and unquenched $N_f$ = 2 simulations \cite{ETMC_B_BSM}. }
\label{fig:Bi_comp}
\end{figure}

\section{Conclusions}
$B_{\rm{K}}$ computation on the lattice has already entered the era  of precision measurements with fully 
unquenched simulations. A number of  
lattice collaborations using various lattice regularizations provide results that are in nice agreement among
themselves. The total estimated uncertainty is less than $4\%$; as a consequence the lattice calculations are not 
responsible any more for the main source of uncertainty in the $\epsilon_{\rm{K}}$ equation (Eq.~\ref{eps_final}).
The comparison of  $B_{\rm{K}}$ results produced with $N_f=2+1$ and  $N_f=2$ unquenched simulations 
indicate that the systematic error due to the quenching of the strange quark is smaller than
other systematic uncertianties.   
The variety of fermion regularisations, today in use, offer a very good control over the contamination of wrong 
chirality operator mixing in the renormalisation of the $\Delta S=2$ four-fermion operator.  
Multiplicative renormalization is guaranteed  with  discretization errors of 
$O(a^2)$. Many  collaborations use non-perturbative methods for the operator renormalization leading to a good  
 control  over an important source of systematic uncertainty. 
However the operator renormalisation still representes  about half of the systematic uncertainties.
Another source of systematic error comes from the fit extrapolation to the physical point; 
both the choice of the fitting function and the relatively large value of the lightest quark mass actually used in the 
simulations for $B_{\rm{K}}$ inevitably inject a non-negligible systematic uncertainty into the final result.  
Of course simulations at  finer lattice spacing in the near future  would make smaller the systematics due to 
lattice artifacts  making  more reliable the continuum limit extrapolation itself as well as the evaluation 
of  renormalization constants. 

Recently a  $N_f=2$ unquenched computation of the matrix elements 
of the full $\Delta S=2$ operator basis emerging from  box diagrams in models beyond the Standard Model has 
been presented by the ETM Collaboration using three values of the lattice spacing. 
The continuum limit extrapolated results  
offer the opportunity of providing constraints on the input parameters of  a class of models BSM.

\vspace*{-0.6cm}
\Acknowledgements
I thank the CKM2010 Organiser Committee for the hospitality and financial support. I wish to thank V. Lubicz and G.C. Rossi for carefully reading  the manuscript and helpful comments. Discussions with G. Martinelli 
and F. Mescia are gratefully acknowledged.

\end{document}